\documentclass{ifacconf}
  \usepackage{graphicx}
  \usepackage{psfrag}
  \usepackage{lipsum}
  \usepackage{xcolor}
  \usepackage{natbib} 
  \usepackage{caption}
    \usepackage{subcaption}

  \usepackage{breqn}
  \usepackage{mathtools}

  \usepackage{amsmath}
  \usepackage{amsfonts}
  \usepackage{enumitem}

  \newcommand\ar{\mathrm{A}}
  \newcommand\dr{\mathrm{D}}
  
  \newcommand\vo{\mathcal{V}}

  \newcommand\go{\mathcal{G}}
  
  \newcommand\gd{\mathcal{G}^{\prime}_k}
  \newcommand\eo{\mathcal{E}}
  \newcommand\ea{\mathcal{E}^{\mathrm{A}}_k}

  \newcommand\ed{\mathcal{E}^{\mathrm{D}}_k}

  \newcommand\ua{U^{\mathrm{A}}_l}
  \newcommand\ud{U^{\mathrm{D}}_l}

  \newcommand\ba{\beta^{\mathrm{A}}}
  \newcommand\bi{\beta^{\mathrm{D}}}
  
  \newcommand\ra{\rho^{\mathrm{A}}}
  \newcommand\ri{\rho^{\mathrm{D}}}
  \newcommand\ka{\kappa^{\mathrm{A}}}
  \newcommand\ki{\kappa^{\mathrm{D}}}

\makeatletter
\newcommand\blfootnote[1]{%
  \begingroup
  \renewcommand\thefootnote{}\footnote{#1}%
  \addtocounter{footnote}{-1}%
  \endgroup
}

\usepackage{multirow}

\begin{document}
\begin{frontmatter}
\title{\LARGE \bf Two-Player Incomplete Games of Resilient Multiagent Systems}
\author[fifth,first]{Yurid E. Nugraha}
\author[first]{Tomohisa Hayakawa}
\author[third]{Hideaki Ishii}
\author[second]{Ahmet Cetinkaya}
\author[fourth]{Quanyan Zhu}

\address[fifth]{Department of  
Electrical Engineering, Sepuluh Nopember Institute of Technology, Surabaya 60111, Indonesia. (yurid@its.ac.id)}
\address[first]{Department of  
Systems and Control Engineering, Tokyo Institute of Technology, Tokyo
152-8552, Japan.  (hayakawa@sc.e.titech.ac.jp)}
\address[third]{Department of Computer Science, Tokyo Insitute of Technology, Yokohama 226-8502,  Japan. (ishii@c.titech.ac.jp)}
\address[second]{Department of Functional Control Systems, Shibaura Institute of Technology, Tokyo
135-8548, Japan. (ahmet@shibaura-it.ac.jp)}
\address[fourth]{Department of Electrical and Computer Engineering, New York University, Brooklyn, NY 11201, USA. (quanyan.zhu@nyu.edu)}

% \thispagestyle{plain}
% \pagestyle{plain}

\begin{abstract}
Evolution of agents' dynamics of multiagent systems under consensus protocol in the face of jamming attacks is discussed, where centralized parties are able to influence the control signals of the agents. In this paper we focus on a game-theoretical approach of multiagent systems where the players have incomplete information on their opponents' strength. We consider repeated games with both simultaneous and sequential player actions where players update their beliefs of each other over time. The effect of the players' optimal strategies according to Bayesian Nash Equilibrium and Perfect Bayesian Equilibrium on agents' consensus is examined. It is shown that an attacker with incomplete knowledge may fail to prevent consensus despite having sufficient resources to do so.
\end{abstract}
\end{frontmatter}

\section{Introduction}
\vspace{-0.5cm}
Jamming attacks\blfootnote{© 2023 the authors. This work has been accepted to IFAC for publication under a Creative Commons Licence CC-BY-NC-ND} in networks are commonly modeled as games between adversaries and agents of networks, e.g., in \citet{jamming1}. In some situations, a player of the game may not exactly know the impact of the players' actions on the utility of its opponents, which may be kept as a private information. We call this lack of information of a player in the game as \textit{incomplete information} \citep{textbook2}. This incomplete information aspect affects how players determine their strategies.

Games with Bayesian probabilities are known to be effective in modelling incomplete/partial information among players. The solution concepts used include Bayesian Nash Equilibrium \citep{bne1} and Perfect Bayesian Equilibrium \citep{pbe1}. These solution concepts are also recently studied in the context of $n$-player networks \citep{netw1,netw2}. In these games, the uninformed players form their \textit{beliefs} of the opponents' private information characterized as \textit{types}.  

There are several game models and solution concepts considering incomplete information of the players under a sequential setting that have been discussed in the literature. For a two-player case, one of the most frequently employed models is signaling games, where an informed player moves first to signal its type to its uninformed opponent \citep{textbook1,sas}. In this formulation, not making actions that indicate its true type may be optimal for a player, depending on the cost. Other variation includes screening games \citep{sign2}, where the uninformed player decides its strategy first without exactly knowing its opponent's type. 

In our previous results, e.g., \cite{arxivYur,eccYur}, we considered a game-theoretical approach for network security problems where there are two centralized players, an attacker and a defender, who strategize on how to attack and defend the network, respectively. There, we considered that the two players under limited resources decide their strategies sequentially. In this paper, we extend the formulation to consider a more realistic scenario where the two players do not know the exact strength of their opponents. It is then natural to consider an incomplete information structure of the game, where a player does not know the exact utilities of its opponent due to its unknown attack/defense strength. 

More specifically, our contribution in this paper is as follows: (i) We consider a two-player game in the context of network security with incomplete information of the players and discuss several solution concepts of the game, namely the Bayesian Nash Equilibrium and the Perfect Bayesian Equilibrium. (ii) We examine how the actions of the rational and myopic players affect the agent states under a consensus protocol in the long term. The players' actions may change over time due to their limited resources and the evolution of their beliefs on the opponents' strengths.

The paper is organized as follows. In Section~\ref{sec2}, we outline the framework for the incomplete information model as well as the attack and defense sequences and energy consumption models of the players.
In Section~\ref{thra}, we focus on the Bayesian Nash Equilibrium where the players execute their strategies simultaneously. We consider the case where the players make their actions sequentially in Section~\ref{thrf}. We then provide numerical examples in Section~\ref{sec5} and conclude the paper in Section~\ref{sec6}.

\vspace{-0.1cm}
\section{Problem Formulation}\label{sec2}
\vspace{-0.3cm}
 We explore a multiagent system of $n$ agents communicating to each other in discrete time. The network topology is described by an undirected and connected graph $\go=(\mathcal{V},\mathcal{E})$. It consists of the set $\mathcal{V}$ of vertices representing the agents and the set $\mathcal{E} \subseteq \mathcal{V} \times \mathcal{V}$ of edges representing the communication links. Each agent $i$ has the scalar state $x_i$ following the consensus update rule at time $k \in \mathbb{N}_0$ given by
    \begin{align} 
        x_i[k+1]&=x_i[k]+u_i[k], \label{state} \\
        u_i[k]&=\sum_{j \in {\mathcal{N}_{i}}[k]} a_{ij}(x_j[k]-x_i[k]), \label{state2}
    \end{align}
where $x[0]=x_0$, $a_{ij}>0$, $\sum_{j=1, j \neq i}^n a_{ij} < 1$, and $\mathcal{N}_i[k]$ denotes the set of agents that can communicate with agent $i$ at time $k  \in \mathbb{N}_0$. This set may change due to the attacks. Under normal operation without any attacks, it is known that all agents converge to the same state \citep{FB-LNS}.

A two-player game between the attacker and the defender is considered in terms of the communication among the agents. The attacker is capable to block the communication by jamming some targeted edges and therefore delay (or completely prevent) the consensus among the agents. These jamming attacks (if successful) are represented by the removal of edges in $\mathcal{G}$. In response to the actions of the attacker, the defender tries to recover the communication by allocating resources in some edges to rebuild those edges under the attacks. Specifically, the defender may ask agent $i$ to send stronger signals to \textit{some} of its neighbors. These strong communication signals make the attacks ineffective. That is, if agent $i$ uses strong communication signals in some edges, then the attacks on those edges will be unsuccessful, i.e., the agents will be able to communicate over those edges. However, by using strong communication signals, agents consume more resources.

\subsection{Attack-Communication Sequence}
\vspace{-0.2cm}
At each time $k$, the attacker decides to attack some edges whereas the defender decides the edges that use strong communication signals. Specifically, at time $k$ the attacker attacks $\go$ by deleting the edges $\ea\subseteq \eo$, whereas the defender chooses the edges to be used with strong communication signals $\ed \subseteq \eo$. As mentioned earlier, the attacker is not able to break the communication with strong signals by the agents. As a consequence, the network changes from $\go$ to $\gd:=(\vo,(\eo \setminus \ea) \cup \ed)$. The agents then communicate to their neighbors $\mathcal{N}_i[k]$ based on this resulting graph $\gd$.

\begin{figure}
        \centering
        \psfrag{e1}{{$\go$}}
        \psfrag{e2}{{$\go^{\prime}_k$}}
        \includegraphics[scale=0.15]{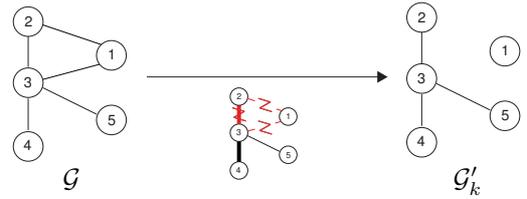}
        %\vspace{-0.3cm}
        \caption{Example of graphs $\go$ and $\go^{\prime}_k$. In the middle graph, the attacked edges are shown in red. The edges used with weak and strong communication signals are illustrated with normal and bold lines, respectively.}
        \label{fig:1}
    \end{figure}

Fig.~\ref{fig:1} illustrates graphs affected by the attacks and communication activities. The original network $\go$ is shown in the left. Then, the attacker attacks edges $\ea=\{(1,2),(1,3),(2,3)\}$, whereas the defender requests agents 2,3, and 4 to communicate strongly through edges $\ed=\{(2,3),(3,4)\}$, and the communication remains weak in other edges $\{(1,2),(1,3),(3,5)\}$. As a result, the agents communicate through edges $\eo^{\prime}_k=\{(2,3),(3,4),(3,5)\}$, shown in the right graph $\gd$.
It is clear that the attacker's actions may not be successful, e.g., the attacker's attack on $(2,3)$ fails due to the strong communication signals by the defender.

  \vspace{-0.2cm}
  \subsection{Resource Constraints}
  \vspace{-0.2cm}
    By attacking (resp., communicating), the attacker (resp., the defender) allocates its resources to some of the edges. These resource allocation actions are affected by the constraints on the resource availability, which is assumed to increase linearly in time. We assume that the resources allocated by the players are proportional to the number of planned attacked/communicated edges. Here, the attacks on $\ea$ cost $\ba>0$ resource per edge. The total resource used by the attacker is constrained as
    \begin{align} 
    \sum_{m=0}^{k} \ba|\eo^{\ar}_m| \leq \ka + \ra k \label{en.a}
    \end{align} 
     for any time $k$, where $\kappa^{\mathrm{A}}\geq \ra>0$ and $\ba>0$. This inequality implies that the total resource spent by the attacker cannot exceed the available resource characterized by the initial resource $\kappa^{\mathrm{A}}$ and the supply rate $\rho^{\mathrm{A}}$. The condition $\ka \geq \ra$ allows the attacker to have at least the same attack ability at time $k=0$, which is important for consensus as discussed later. This resource constraint upper-bounds the number of edges that the attacker can attack. See also \citep{en1,en2} and the references therein for related attack models.
    
    The resource constraint of the defender, which is similar to (\ref{en.a}), is given by
    \begin{equation} \label{en.d0}
        \sum_{m=0}^{k} (\hat{\beta}^{\dr} |\eo^{\dr}_m| + \overline{\beta}^{\dr} |\overline{\eo}^{\dr}_m|) \leq \hat{\kappa}^{\dr} + \hat{\rho}^{\dr} k
    \end{equation}
     with $\hat{\rho}^{\dr}>\overline{\beta}^{\dr}|\eo|>0$, $\hat{\kappa}^{\dr}>\overline{\beta}^{\dr}|\eo|>0$, and $\hat{\beta}^{\dr}>\overline{\beta}^{\dr}>0$. We denote by $\overline{\beta}^{\dr}$ and $\hat{\beta}^{\dr}$ the unit cost of communicating with weak and strong signals respectively, whereas $\overline{\eo}^{\dr}_m$ is used to denote the number of edges used with weak communication signals. The inequality $\hat{\beta}^{\dr}>\overline{\beta}^{\dr}$ implies the fact that the consumed resources with the strong signals are larger than those with the weak signals, whereas $\hat{\kappa}^{\dr},\hat{\rho}^{\dr}>\overline{\beta}^{\dr}|\eo|$ implies that the agents are able to communicate with weak signals in all edges at all times.
     
     Since $\overline{\eo}^{\dr}_m=\eo\setminus \eo^{\dr}_m$, by supposing $\hat{\rho}^{\dr}=\rho^{\dr}+\overline{\beta}^{\dr}|\eo|$, $\hat{\kappa}^{\dr}={\kappa}^{\dr}+\overline{\beta}^{\dr}|\eo|$, and $\hat{\beta}^{\dr}=\beta^{\dr}+\overline{\beta}^{\dr}|\eo|$, we can rewrite (\ref{en.d0}) as
    \begin{align*} 
        \sum_{m=0}^{k} (\hat{\beta}^{\dr}-&\overline{\beta}^{\dr}) |\eo^{\dr}_m| + \overline{\beta}^{\dr}|\eo|(k+1) \\ &\leq \ki + \ri k + \overline{\beta}^{\dr}|\eo| (k+1)
    \end{align*}
    and hence
    \begin{align} \label{en.d}
        \sum_{m=0}^{k} \bi |\eo^{\dr}_m| &\leq \ki + \ri k.
    \end{align}
    For the rest of the paper, for simplicity we call $\bi$ and $\rho^{\dr}$ as the unit cost and the supply rate of the defender, respectively. 

\subsection{Games with Incomplete Information}
\vspace{-0.2cm}
We then continue by explaining the structure of the game between the attacker and the defender. We first make several assumptions regarding each player's knowledge of the other player's parameters. 

We consider that the attacker does not know the cost $\bi$ of the defender. Similarly, the cost $\ba$ of the attacker is also not known by the defender. For simplicity, other resource parameters $\ka$, $\ki$, $\ra$, and $\ri$ as well as the agent states at time $k$ are known by both players.

Throughout this paper, we suppose that both players are strategic, i.e., they execute the strategies associated with the equilibrium. We later discuss how these strategies affect the agents' dynamics in the long term. In addition, we assume that both players are aware of the limitation in the knowledge of the other player. For example, the defender knows that the attacker does not have the exact value of $\bi$.

Different from games with complete information of the players, in games with incomplete information players have to guess the \textit{type} of their opponents based on a certain probability distributions called \textit{beliefs} \citep{textbook2,textbook1}. Since a game is played every time $k$, in this paper we suppose that the players update their beliefs over time, as explained later.
    
In this game we suppose that the types of the attacker unknown by the defender are determined by the unit cost $\ba$. We denote those types as $\theta^{\ar}\in\Theta^{\ar}=\{\beta^{\ar}_1,\beta^{\ar}_2\}$ where $\beta^{\ar}_1$, $\beta^{\ar}_2$ denote the possible unit cost values of the attacker and satisfy $\beta^{\ar}_1<\beta^{\ar}_2$. The defender's belief of the attacker's type at time $k$ is denoted as $0 \leq \mu^{\dr}_k(\theta^{\ar})\leq 1$, $\theta^{\ar} \in \Theta^{\ar}$. For simplicity, we assume only two types $\beta^{\ar}_1$ and $\beta^{\ar}_2$. We denote the beliefs of the types $\mu^{\dr}_k(\beta^{\ar}_1)$ and $\mu^{\dr}_k(\beta^{\ar}_2)$ as $\mu^{\dr,1}_{k}$ and $\mu^{\dr,2}_{k}$, respectively, with $\mu^{\dr,1}_{k}+\mu^{\dr,2}_{k}=1$.
    
Similarly, the defender's types are determined by its unit cost $\beta^{\dr}$. We thus denote its types as $\theta^{\dr} \in\Theta^{\dr}=\{\beta^{\dr}_1,{\beta}^{\dr}_2\}$, with $\beta^{\dr}_1<\beta^{\dr}_2$. The attacker's belief of the defender's type at time $k$ is denoted as $0\leq \mu^{\ar}_k(\theta^{\dr})\leq 1, \theta^{\dr}\in\Theta^{\dr}$. Denote the attacker's beliefs $\mu^{\ar}_k(\beta^{\dr}_1)$ and $\mu^{\ar}_k(\beta^{\dr}_2)$ as $\mu^{\ar,1}_{k}$ and $\mu^{\ar,2}_{k}$, respectively, with $\mu^{\ar,1}_{k}+\mu^{\ar,2}_{k}=1$. These types $\theta^{\ar}$ and ${\theta}^{\dr}$ do not change over time $k$.
    
In this paper, we focus on the incompleteness of the players' information. We discuss two cases:
\begin{enumerate}[label=(Case \arabic*), leftmargin=*]
    \item The attacker sends jamming signals simultaneously as the defender asks agents to send communication signals.
    \item The attacker sends jamming signals before the defender decides how the agents should communicate (both actions are still considered to happen at time $k$).
\end{enumerate}

In Case~1, the defender does not observe which edges are attacked and thus does \textit{not} have \textit{perfect} information of the attacks. We then consider the Bayesian Nash Equilibrium (BNE) as a suitable solution concept in this case. We discuss this formulation in more detail in Section~\ref{thra}.

On the other hand, since in Case~2 the attacks and the communications occur sequentially, the defender is aware of the attacks in $\eo$ and decides its strategy based on the attacks. In this case we consider the Perfect Bayesian Equilibrium (PBE) for the interaction between the attacker and the defender, which will be discussed in Section~\ref{thrf}. For simplicity, in Case~2 we consider only one uninformed player, i.e., only one player has incomplete knowledge, whereas the other player knows the type of the opponent.    

\subsection{Agent State Difference and Utility Functions Design}
\vspace{-0.2cm}
    In our problem setting, the players also consider the effects of their actions on the agent states when attacking/communicating. Specifically, the attacker wants to make the difference among the agent states $x_i[k]$ as large as possible, whereas the defender attempts to keep this difference as small as possible. To this end, we specify the sum of the agents' state differences $z_k$ of time $k$ as
    \begin{align}
    z_k(\ea,\ed):= x^{\mathrm{T}}[k+1] L_{\mathrm{c}} x[k+1], \label{z}
    \end{align}
    with $L_{\mathrm{c}}\in \mathbb{R}^{n\times n}$ being the Laplacian matrix of the complete graph with $n$ agents. The choices of edges $(\ea,\ed)$ will affect $x[k+1]$ in (\ref{state}) and in turn the value of $z_k$.
    
    The utility functions of the players at the $k$th game (played at time $k$) for both BNE and PBE cases given the players' types $\theta^{\ar}$ and $\theta^{\dr}$ are defined by
  \begin{align}
    \hat{u}^{\ar}_k(\theta^{\dr},\ea,\ed)  :=& z_k+\theta^{\dr}|\eo^{\mathrm{D}}_k|-\ba|\ea|, \label{ua}
    \\
    \hat{u}^{\dr}_k (\theta^{\ar},\ea,\ed) :=& -z_k-{\beta}^{\dr}|\eo^{\mathrm{D}}_k|+\theta^{\ar}|\ea| \label{ud},
    \end{align}
    which are to be maximized by the players. These functions consider both players' costs of attacking/communicating strongly as well as the effect of their actions on the agents' state difference. Note that in (\ref{ua}), (\ref{ud}) we take account of the players' types $\theta^{\ar}$, $\theta^{\dr}$ in the form of the cost terms ${\theta}^{\dr}|\eo^{\mathrm{D}}_k|$ and ${\theta}^{\ar}|\ea|$. Recall that the players may waste their resources by allocating them inefficiently, e.g., the attacker may attempt to attack an edge over which agents use strong communication signals. We then can see from (\ref{ua}) (resp., (\ref{ud})) that the inefficient actions by the attacker (resp., the defender) will not increase (resp., decrease) the values of $z_k$ but affect the utilities negatively by increasing the values of $|\ea|$ (resp., $|\eo^{\mathrm{D}}_k|$).
    
    Due to the uncertainty in the players' various types $\theta^{\ar}$, $\theta^{\dr}$, the exact values of (\ref{ua}) and (\ref{ud}) are not known by the players. To find the equilibria, we thus evaluate the possible utilities across all types of the players to obtain the expected utilities, which are represented by
    \begin{align}
        U^{\ar}_k(\ea,\ed)&= \sum_{\theta^{\dr}\in \Theta^{\dr}} \mu_k^{\ar} (\theta^{\dr}) \hat{u}^{\ar}_k(\theta^{\dr},\ea,\ed), \label{ex.a}
     \\
    %   \end{align}
    % \begin{align}
        U^{\dr}_k(\ea,\ed)&= \sum_{\theta^{\ar} \in \Theta^{\ar}} \mu_k^{\dr} (\theta^{\ar}) \hat{u}^{\dr}_k(\theta^{\ar},\ea,\ed). \label{ex.d}
    \end{align}
    
     These expected utilities are to be maximized by the players at every time step $k$.
     
         \subsection{Belief Update Design}
         \vspace{-0.2cm}
    In this game, we suppose that the players update their beliefs over time by considering their previous actions. Since the beliefs affect the players' optimal strategies, it is important for the players to have good beliefs of their opponents' types to use their resources efficiently. 
    
    We design the belief update system as follows. From the resource constraints of the players specified in (\ref{en.a}) and (\ref{en.d}) above, at time $k$ the players obtain their \textit{predicted costs} $\tilde{\beta}_k^{\ar}$ and $\tilde{\beta}^{\dr}_k$ (not necessarily in $\Theta^{\ar}$ or $\Theta^{\dr}$) specified by
    \begin{align}
        \tilde{\beta}^{\ar}_k=\frac{\ka+\ra (k-1)}{\sum_{m=0}^{k-1}|\eo^{\ar}_m|}, \quad \tilde{\beta}^{\dr}_k=\frac{\kappa^{\dr}+\rho^{\dr} (k-1)}{\sum_{m=0}^{k-1}|\eo^{\dr}_m|},
    \end{align}
    which are updated at every time $k$. 
    
    These predicted costs $\tilde{\beta}_k^{\ar}$ and $\tilde{\beta}^{\dr}_k$ are obtained by supposing that from time 0 to time $k-1$
    the players' attack/communicate actions are known to each other. This is a common assumption and not difficult to realize in practice. The players predict their opponents' types by utilizing these predicted costs. For example, if the value of $\tilde{\beta}_k^{\ar}$ is close enough to $\beta^{\ar}_2$, then the defender has a larger $\mu^{\dr,2}_k$ compared to $\mu^{\dr,1}_k$.
    
    We first address the situation where there is no previous attack (resp., no previous communication with strong signals), i.e., $\sum_{m=0}^{k-1}|\eo^{\ar}_m|= 0$ (resp., $\sum_{m=0}^{k-1}|\eo^{\dr}_m|= 0$). In this case, the attacker (resp., the defender) assigns equal beliefs $\mu_{k}^{\ar,1}=\mu_{k}^{\ar,2}=0.5$ (resp., $\mu_{k}^{\dr,1}=\mu_{k}^{\dr,2}=0.5$).
    
    Furthermore, since the players' types do not change over time, if a player is sure of its opponent's type $\theta^{\ar}$ or $\theta^{\dr}$, i.e., $\mu_{k^{\prime}}^{\dr}(\theta^{\ar})=1$ or $\mu_{k^{\prime}}^{\ar}(\theta^{\dr})=1$ at time $k^{\prime}$, then the beliefs of the subsequent time $k>k^{\prime}$ do not change.
    
    Assuming $\sum_{m=0}^{k-1}|\eo^{\ar}_m| \ne 0$ and $\mu_{k^{\prime}}^{\ar}(\theta^{\ar})\ne 1$ for any type $\theta^{\ar}$ and time $k^{\prime}<k$, the attacker's belief of the defender's type at time $k$ is
    \begin{align}\label{bel_a}
        \mu_k^{\ar,1}=1-\mu_k^{\ar,2}= \left\{
	\begin{array}{ll}
		\alpha, & \mbox{if } \tilde{\beta}^{\dr}_k={\beta}^{\dr}_2 \\
		1, & \mbox{if } \tilde{\beta}^{\dr}_k<{\beta}^{\dr}_2 \\
		\frac{0.5(\tilde{\beta}^{\dr}_k-{\beta}^{\dr}_2)}{\tilde{\beta}^{\dr}_k}, & \mbox{otherwise.}
	\end{array}
\right.
    \end{align}
    for the attacker, with $\alpha<0.5$. Similarly, assuming $\sum_{m=0}^{k-1}|\eo^{\dr}_m| \ne 0$ and $\mu_{k^{\prime}}^{\dr}(\theta^{\dr})\ne 1$ for any $k^{\prime}<k$ the defender predicts the attacker's type with the belief updated as 
    \begin{align} \label{bel_d}
        \mu_k^{\dr,1}=1-\mu_k^{\dr,2}= \left\{	\begin{array}{ll}
		\alpha, & \mbox{if } \tilde{\beta}^{\ar}_k={\beta}^{\ar}_2 \\
		1, & \mbox{if } \tilde{\beta}^{\ar}_k<{\beta}^{\ar}_2 \\
		\frac{0.5(\tilde{\beta}^{\ar}_k-{\beta}^{\ar}_2)}{\tilde{\beta}^{\ar}_k}, & \mbox{otherwise.}
	\end{array}
	\right.
    \end{align}
    
    From (\ref{bel_a}) (resp., (\ref{bel_d})), we see that if the predicted cost at time $k$ is smaller than the larger cost $\beta^{\dr}_2$ (resp., $\beta^{\ar}_2$), then the attacker (resp., the defender) is sure that its opponent has type $\beta^{\dr}_1$ (resp., $\beta^{\ar}_1$), since with costs $\beta^{\dr}_2$ (resp., $\beta^{\ar}_2$) it is impossible to communicate strongly $\sum_{m=0}^{k-1} |\eo^{\dr}_m|$ (resp., to attack $\sum_{m=0}^{k-1} |\eo^{\ar}_m|$) number of edges. On the other hand, if the predicted cost is very large (much larger than $\beta^{\ar}_2$ or $\beta^{\dr}_2$), then the players become less sure of the type of the opponent.
% %     \vspace{-0.1cm}
    \section{Bayesian Nash Equilibrium under Incomplete and Imperfect information} \label{thra}
    \vspace{-0.3cm}
    In this section, we consider Case 1 characterized in Section~\ref{sec2}, where both players execute their strategies simultaneously. We first discuss our results in terms of the equilibria and the resulting optimal strategies. Then, we analyze the effects of those optimal strategies on consensus and agents' dynamics.

    % can obtain the calculate the past resource of players with We consider a middle value of cost between two types of each player $\tilde{\beta}^{\ar}=\frac{\beta^{\ar}_1+ \beta^{\ar}_2}{2}$ for the attacker and $\tilde{\beta}^{\dr}=\frac{\beta^{\dr}_1+ \beta^{\dr}_2}{2}$ for the defender. We then calculate the past If a player consume all of its resource  Specifically, if an agent 
    \subsection{BNE Analysis}
    \vspace{-0.2cm}
    Here we state several results of this formulation. Since it is difficult to directly derive the general results for any number of agents and graph structures, we begin by stating the results for special cases.
    
    We first consider a case where the defender does not have enough resources for communicating even over one edge with strong signals. Note that even though the attacker does not know the exact value of $\bi$, the edges used to communicate strongly in the past $\eo^{\dr}_{k^{\prime}}$, $k^{\prime}<k$, are known by both players. It then follows that if the defender with $\bi=\beta^{\dr}_1$ would not have enough resource to communicate strongly, then the attacker will try to attack some edges as stated below.
    
    \begin{lem}\label{basic}
    In the game under Case~1, suppose that $\ki+\ri k<\sum_{m=0}^{k-1}{\beta}^{\dr}_1|\eo^{\dr}_m|+\beta^{\dr}_1$ is satisfied at time $k$. Then, $\ea=\emptyset$ is not optimal for the attacker at time $k$.
    \end{lem}
    \begin{pf}
    If $\sum_{m=0}^{k-1}\beta^{\dr}|\eo^{\dr}_m|+\beta^{\dr}> \ki+\ri k$ is satisfied, then the defender is not able to communicate strongly with any edge due to limited resources as constrained in (\ref{en.d}). Since the defender's possible cost satisfies $\beta^{\dr}_1<\beta^{\dr}_2$, we have $\ki+\ri k<\sum_{m=0}^{k-1}{\beta}^{\dr}_1|\eo^{\dr}_m|+\beta^{\dr}_1<\sum_{m=0}^{k-1}{\beta}^{\dr}_2|\eo^{\dr}_m|+\beta^{\dr}_2$. Thus, with the defender's cost either $\beta^{\dr}_1$ or $\beta^{\dr}_2$, the attacker knows that there is no strong communication and thus will try to attack as many edges as possible. \hfill $\square$
    \end{pf}
    
    Similarly, for the attacker's case, we can show that if the attacker with $\ba=\beta^{\ar}_1$ would not have enough resource to attack any edge, then the defender will not use any edge with strong signals as stated in Lemma~\ref{basic2} below.
    
    \begin{lem}\label{basic2}
    In the game under Case 1, suppose that $\ka+\ra k - \sum_{m=0}^{k-1}\beta^{\ar}_1|\eo^{\ar}_m|< \beta^{\ar}_1$ is satisfied at time $k$. Then, $\eo^{\dr}_k=\emptyset$ is an optimal strategy for the defender at time $k$.
    \end{lem}
    % \begin{pf}
    % The proof is similar to the proof of Lemma~\ref{basic}. \hfill $\square$
    % \end{pf}
    
    Let us concentrate on the BNE of the two-agent case with one edge, with $a_{12}=a_{21}$ in (\ref{state}). In this case, there is only one possible change of $z_k$ by the action of attacks/communications of the only edge $(1,2)$. We then denote $z_k$ with and without communication using edge $(1,2)$ at time $k$, respectively, as $z^0_k:=(x_1[k]-x_2[k])^2$ and $z^1_k:=(x_1[k+1]-x_2[k+1])^2=((1-2a_{12})(x_1[k]-x_2[k]))^2$. We will consider the difference $\tilde{z}_k:=z^0_k-z^1_k=4a_{12}(1-a_{12})(x_1[k]-x_2[k])^2$ as well as the players' costs in the results below.
    
    In Lemma~\ref{2case1}, we state that if the agent states are close enough, then not communicating is an optimal strategy for the defender.
    
    \begin{lem}\label{2case1}
    In the game under Case 1, suppose that $\beta^{\dr}_1>\tilde{z}_k.$ Then, $\ed=\emptyset$ is an optimal strategy for the defender regardless of the attacker's actions.
    \end{lem}
    
    \begin{pf}
    In the two-agent case, the defender's utility is at most $\hat{u}^{\dr}_k=-z^1_k-\beta^{\dr}_1$ when there is communication with weak signals without attack and the type of the defender is $\beta^{\dr}_1<\beta^{\dr}_2$. Thus, if $z^1_k+\beta^{\dr}_1>z^0_k$, then the defender will always prefer not to communicate with strong signals. \hfill $\square$
    \end{pf}
    
    Similar characteristics also exist for the attacker, which can be shown similarly to Lemma~\ref{2case1}.
    \begin{lem}\label{2case2}
    In the game under Case 1, suppose that $\beta^{\dr}_1>\tilde{z}_k.$ Then, $\ea=\emptyset$ is optimal for the attacker regardless of the defender's actions and costs.
    \end{lem}
    % \begin{pf}
    % The proof is similar to the proof of Lemma~\ref{2case1}. \hfill $\square$
    % \end{pf}
    
    From Lemmas~\ref{2case1} and \ref{2case2}, we see that if the agent states are close enough at time $k$, then both players will prefer to do nothing. Since the difference among the agent states does not increase over time \citep{FB-LNS}, there will be no action in subsequent times as stated in the next result.
    
    \begin{cor}\label{cor1}
    Suppose that $\min\{\beta^{\dr}_1,\beta^{\ar}_1\}>\tilde{z}_{k^{\prime}}$ at time $k^{\prime}$. Then, in the game under Case 1 at any time $k\geq k^{\prime}$, there is no attack nor communication with strong signals.
    \end{cor}
    
    \subsection{Effect of BNE on Consensus}
    \vspace{-0.2cm}
    Here, we discuss how the actions of strategic players acting simultaneously following BNE affect the agents' dynamics. Firstly, we address the two-agent case. Specifically, from Corollary~\ref{cor1}, we can infer that if two agents have their initial states close enough, i.e., $(x_1[0]-x_2[0])^2<\beta^{\ar}_1$, then consensus will be achieved at infinite time. The following lemma addresses the general case of $n$ agents.
    
    \begin{lem}\label{cons1}
    In the game under Case~1, consensus will be achieved if there exists time $k^{\prime}$ such that $z_{k^{\prime}}<\beta^{\ar}_1$.
    \end{lem}
    \begin{pf}
     At time $k^{\prime}$, the attacker's utility is $\hat{u}^{\ar}_{k^{\prime}}(\eo^{\ar}_{k^{\prime}}\ne \emptyset)=z_{k^{\prime}+1}-\ba|\eo^{\ar}_{k^{\prime}}|+\theta^{\dr}|\eo^{\dr}_{k^{\prime}}|$ if it chooses to attack. Since $z_{k^{\prime}+1}\leq z_{k^{\prime}}<\beta^{\ar}_1$, the attacker will receive negative utility by attacking. This is always worse than the utility without attack $\hat{u}^{\ar}_{k^{\prime}}(\eo^{\ar}_{k^{\prime}}\ne \emptyset)>0$, since $z_{k^{\prime}}>0$. It then follows that the attacker does not attack any edge, and since $z_k$ does not increase with the consensus protocol (\ref{state2}), there will be no attack in the future time $k>k^{\prime}$, and as a result consensus is achieved. \hfill $\square$
    \end{pf}

    \section{Perfect Bayesian Equilibrium under Incomplete Information}\label{thrf}
    \vspace{-0.3cm}
    
    \begin{figure}
    \begin{subfigure}[b]{0.5\textwidth}
        %\centering
        \hspace{1cm}
        \psfrag{A1}{\scriptsize $\theta^{\dr}=\beta^{\dr}_1$}
        \psfrag{A2}{\scriptsize $\ed$}
        \psfrag{A3}{\scriptsize $\theta^{\dr}=\beta^{\dr}_2$}
        \psfrag{A4}{\scriptsize $\ea$}
        \includegraphics[scale=0.5]{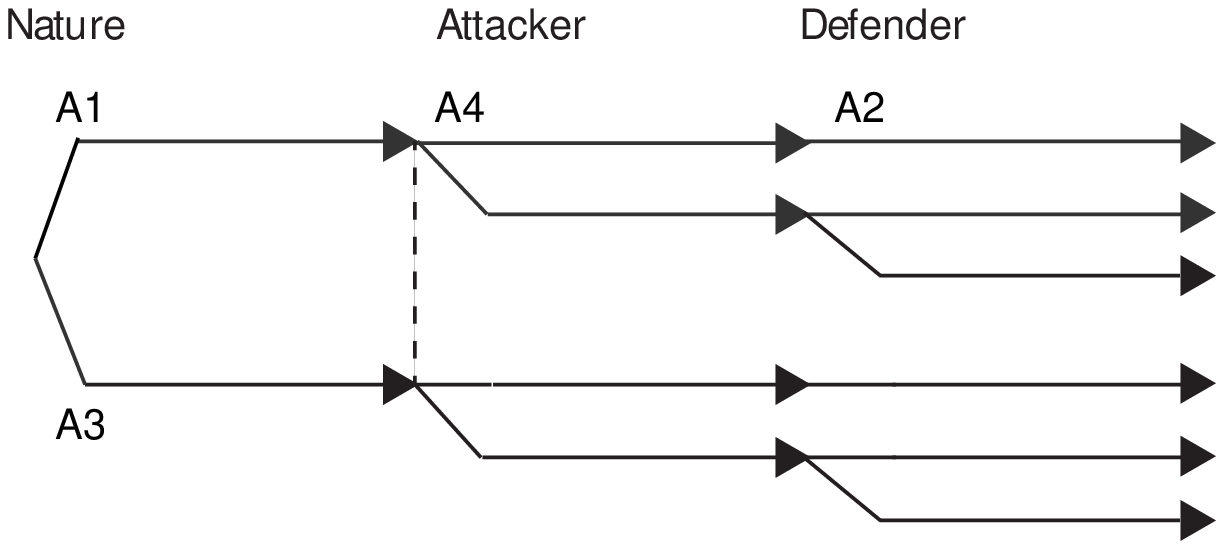}
        \subcaption{A screening game structure}
        \vspace{0.3cm}
        \label{screen1}
    \end{subfigure}
    \begin{subfigure}[b]{0.5\textwidth}
        \hspace{1cm}
        \psfrag{A1}{\scriptsize $\theta^{\ar}=\beta^{\ar}_1$}
        \psfrag{A2}{\scriptsize $\ed$}
        \psfrag{A3}{\scriptsize $\theta^{\ar}=\beta^{\ar}_2$}
        \psfrag{A4}{\scriptsize $\ea$}
        \includegraphics[scale=0.5]{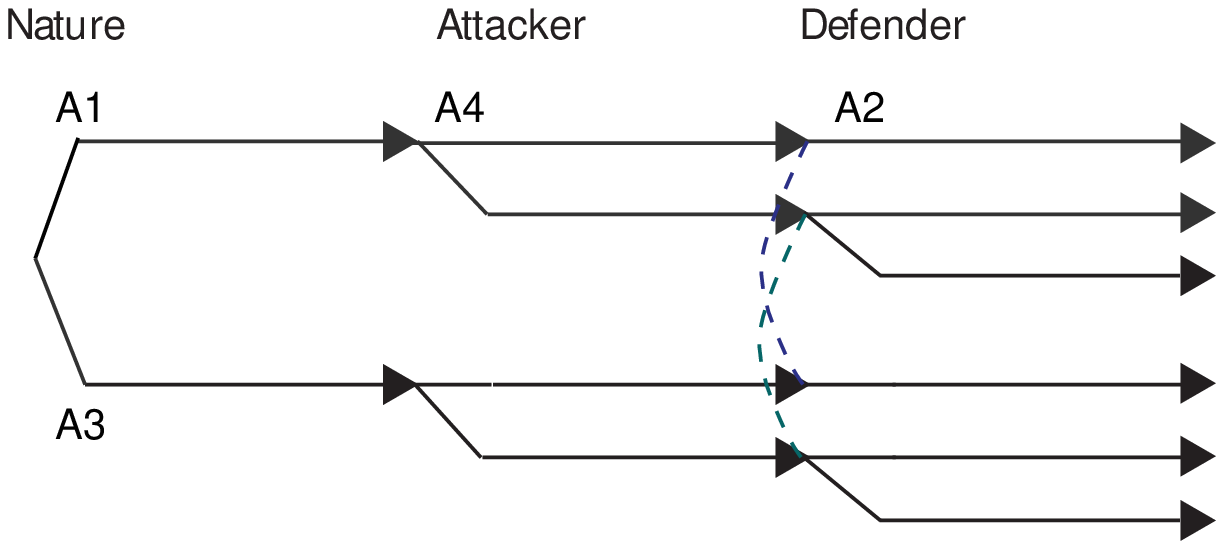}
        \subcaption{A signaling game structure}
        \vspace{0.0cm}
    \end{subfigure}
         \caption{A screening (resp., signaling) game structure where the dashed line illustrates the incomplete information by the attacker (resp., the defender), respectively, i.e., the attacker (resp., the defender) does not know where it is on the game tree}
        \label{screen2}
\end{figure}

    We now consider Case~2 characterized in Section~\ref{sec2}, i.e., the situation where the attacker acts first followed by the defender (all actions are still made at the same time $k$). Therefore, the defender observes which edges are jammed by the attacker among $\eo$ before making its decision.
    
    Similar to the discussion on Case~1 in Section~\ref{thra}, in sequential games with incomplete information we consider that the players admit certain types $\theta^{\ar}$ and $\theta^{\dr}$. Here we also assume that the players' beliefs are updated over time given their knowledge of the past players' strategies before time $k$. Specifically, we consider the belief update rules over time $k$ in (\ref{bel_a}) and (\ref{bel_d}).
    
    However, in Case~2, since the defender observes the attacker's move at the same $k$, its belief is also updated at the same $k$ based on the attacker's action $\ea$. We thus denote the defender's belief as $\mu^{\dr}_k(\theta^{\ar}|\ea)$ updated at the same $k$ according to the Bayesian rules \citep{textbook2}.
        
    As mentioned in Section~\ref{sec2}, we consider two subcases of Case~2: (Case~2a) The attacker does not know the defender's type, and (Case~2b) the defender does not know the attacker's type. These two subcases will be examined in Sections~\ref{41} and \ref{42}, respectively. The case where both players do not know each other's types requires more analysis and is not investigated in this paper.
    
    \subsection{PBE with Unknown Defender Type}\label{41}
    \vspace{-0.2cm}
    
    In this section we examine the case where the attacker does not know whether the defender's cost is either $\beta^{\dr}_1$ or $\beta^{\dr}_2$. Incomplete information games where the uninformed player makes its action first (in this case, the attacker) are sometimes called \textit{screening games} \citep{sign2}.
    
    The extensive-form structure of this game is illustrated in Fig.~\ref{screen1}. At first, the $\textit{nature}$ decides the type of the defender $\theta^{\dr}$, which is not known by the attacker. In our setting of games played repeatedly, we suppose that the nature always chooses the same type at any $k$. Then the attacker chooses its strategy followed by the defender, who knows its own type as well as the attacker's type. 
    
    Similarly to the discussion of BNE in Section~\ref{thra} above, here we begin by ruling out some cases. Specifically, for no-resource case of the defender, i.e., $\ki+\ri k - \sum_{m=0}^{k-1}(\tilde{\beta}^{\dr}_1|\eo^{\dr}_m|)< \beta^{\dr}_1$, Lemma~\ref{basic} also holds for the PBE case at time $k$. Likewise, for no-resource case of the attacker, Lemma~\ref{basic2} holds.
    
    We then continue to discuss the effects of the knowledge of the attacks on the defender's communication actions. Since we suppose that the defender is aware of the attacks in this case, communicating with strong signals in unattacked edges is never desirable for the defender.
    
    \begin{lem} \label{waste1}
    In the game under Case~2, $\ed$ satisfying $(\ed\setminus\ea)\ne \emptyset$ is not optimal for the defender for any time $k$ and any types and parameters of the players.
    \end{lem}
    \begin{pf}
    From (\ref{ud}), it is clear that if $\tilde{\eo}_k=\ed\setminus\ea\ne \emptyset$ the defender does not obtain additional value of $-z_k$ but obtains negative value of $-\bi|\ed|$. It then follows that for any given $\ea$ and across all types $\Theta^{\ar}$, the expected utility (\ref{ex.d}) satisfies $U^{\dr}_k(\tilde{\eo}_k\ne\emptyset)<U^{\dr}_k(\tilde{\eo}_k=\emptyset)$.\hfill $\square$
    \end{pf}

    % We then discuss the strategies of the attacker. If the attacker has enough resource to attack all edges $\eo$, then the defender may communicate with strong signals to prevent the attack. This results in no-attack action by the attacker, and as a result the defender ends up communicating with weak signals.
    
    We then consider a special case of two-agent communication with one edge $(1,2)$. From Lemma~\ref{waste1}, it is clear that the defender does not use strong signals if the attacker does not attack. If the attacker attacks, i.e., $\ea=\{(1,2)\}$, then the defender will use strong signals if and only if $z_k^0-z^1_k>\theta^{\dr}$. This implies that the defender's decision can differ based on its type. That is, if $\beta^{\dr}_1\leq z_k^0-z^1_k \leq \beta^{\dr}_2$, then the defender chooses to communicate strongly if its type is $\beta^{\dr}_1$ and not if its type is $\beta^{\dr}_2$.

    \begin{prop}\label{screen}
    In the game under Case 2a, for $n=2$, the players' optimal strategies can be classified as follows:
    \begin{enumerate}[label=(\alph*)]
        \item For $\tilde{z}_k>\beta^{\dr}_2$, the attacker attacks if $\mu_k^{\ar,1}<\frac{\beta^{\dr}_2-\beta^{\ar}}{\beta^{\dr}_2-\beta^{\dr}_1}$ and does not otherwise, and the defender uses strong signals regardless of its type.
        \item For $\tilde{z}_k < \beta^{\dr}_1$, the attacker attacks if $\tilde{z}_k>\ba$ and does not otherwise, whereas the defender does not use strong signals in any type.
        \item For $\beta^{\dr}_1 \leq \tilde{z}_k \leq \beta^{\dr}_2$, the defender uses strong signals if its type is $\beta^{\dr}_1$ and does not if its type is $\beta^{\dr}_2$. The attacker attacks if $\mu^{\ar,1}_k>\frac{\tilde{z}_k-\beta^{\ar}}{\tilde{z}_k+\beta^{\dr}_1}$.
    \end{enumerate}
    These optimal strategies cover all cases of the values of $z^0_k$, $z^1_k$, $\beta^{\dr}_1$, and $\beta^{\dr}_2$.
    \end{prop}
    \begin{pf}
    First, in the case of no attack, it is clear that the defender does not use strong signals for both types and thus the expected utility satisfies $U^{\dr}_k(\emptyset,\emptyset)=-U^{\ar}_k(\emptyset,\emptyset)=-z^1_k$. 
    
    We now consider the three cases (a)--(c) based on the value of $\tilde{z}_k$.
    
    Case (a): $\tilde{z}_k>\beta^{\dr}_2$. If there is an attack on edge $(1,2)$, the defender will use strong signals for both types if $z^0_k-z^1_k>\beta^{\dr}_2$. Thus, the expected utility for the attacker is $U^{\ar}_k((1,2),(1,2))=\mu_k^{\ar,1}(z^1_k-\ba+\beta^{\dr}_1)+(1-\mu_k^{\ar,1})(z^1_k-\ba+\beta^{\dr}_2)$. This will then imply that the attacker will decide to attack if $\mu_k^{\ar,1}<\frac{\beta^{\dr}_2-\beta^{\ar}}{\beta^{\dr}_2-\beta^{\dr}_1}$.
    
    Case (b): $\tilde{z}_k < \beta^{\dr}_1$. Here, it is understood that the defender does not use strong signals for both types. The expected utility of the attacker is $U^{\ar}_k((1,2),(1,2))=\mu_k^{\ar,1}(z^0_k-\ba)+(1-\mu_k^{\ar,1})(z^0_k-\ba)=z^0_k-\ba$. Thus, the attacker will choose to attack if $z^0_k-z^1_k>\ba$, regardless of its belief. 
    
    Case (c): $\beta^{\dr}_1 \leq \tilde{z}_k \leq \beta^{\dr}_2$. The defender will use strong signals if its type is $\beta^{\dr}_1$ and not if its type is $\beta^{\dr}_2$. In this case, the expected utility for the attacker if it decides to attack is $U^{\ar}_k((1,2),\ed)=\mu_k^{\ar,1}(z^1_k-\ba+\beta^{\dr}_1)+(1-\mu_k^{\ar,1})(z^0_k-\ba)=\mu^{\ar,1}_k(z^1_k+\beta^{\dr}_1-z^0_k)+z^0_k-\beta^{\ar}$. The attacker will choose to attack if its belief satisfy $\mu^{\ar,1}_k>\frac{z^1_k-z^0_k-\beta^{\ar}}{z^1_k-z^0_k+\beta^{\dr}_1}$.
    $\hfill \square$
    \end{pf}
    
    It is interesting to note from the result above that when $\tilde{z}_k>\beta^{\dr}_2$, the attacker chooses to attack if its belief of high cost of the defender, i.e., $\beta^{\dr}_2$ type, is relatively high.
    
    \subsection{PBE with Unknown Attacker Type}\label{42}
    \vspace{-0.2cm}
    In this subsection we examine the case where the defender, which moves later, does not know whether the attacker's cost is $\beta^{\ar}_1$ or $\beta^{\ar}_2$. The two-player incomplete information games where the uninformed player makes its action last (in this case, the attacker) is commonly known as \textit{signaling games} \citep{textbook2}.
    
    The extensive-form game structure of signaling games is illustrated in Fig.~\ref{screen2}. At first, the nature decides the attacker's type $\theta^{\ar}$, which is not known by the defender (we suppose that the nature always chooses the same type at any $k$). Then the attacker chooses its strategy followed by the defender, who does not know the attacker's type despite knowing its actions.

    Again, here we begin by ruling out some cases regarding the resource limitation. Specifically, Lemmas~\ref{basic} and \ref{basic2} hold for signaling games at time $k$.

    We first characterize the defender's best response given the attacker's action. If there is no attack, from the defender's utility functions in (\ref{ud}) it can be understood that there is no communication with strong signals. On the other hand, if there is an attack, the defender's utility becomes $\ud=-z^1_k+\theta^{\ar}-\bi$ if it chooses to recover and $\ud=-z^0_k+\theta^{\ar}$ if it chooses not to do so. Thus, the defender will use strong signals the edge only if $z^0_k-z^1_k>\bi$. Note that the defender's best response here is not affected by the type of the attacker, unlike the attacker's response in screening games specified above.

    Now we are ready to state the results of the equilibrium in signaling games for $n=2$. In signaling games, there are three types of equilibria: (i) Separating equilibrium: different types have different actions, (ii) Pooling equilibrium: all types have the same actions, and (iii) Semi-separating equilibrium: the types of the players affect the actions of the players in a stochastic way. We discuss those equilibria in this section. 
    
    \begin{prop} \label{sep}
    In the game under Case 2b, for $n=2$, a separating equilibrium exists if one of the following conditions is satisfied:
    \begin{itemize}
        \item $\beta^{\ar}_1\leq \tilde{z}_k \leq \beta^{\ar}_2$ and $\tilde{z}_k\leq\bi$, or
        \item $\beta^{\ar}_1\leq \beta^{\dr} \leq \beta^{\ar}_2$ and $\tilde{z}_k>\bi$.
    \end{itemize}
     The optimal strategies of the players are as follows:
    \begin{itemize}
        \item If the attacker's type is $\beta^{\ar}_1$, then attacking is an optimal strategy. Otherwise, if the attacker's type is $\beta^{\ar}_2$, then not attacking is an optimal strategy.
        \item The defender uses strong signals if $\tilde{z}_k>\bi$, and not otherwise.
    \end{itemize}
    \end{prop}
    \begin{pf}
    From the explanation above, the defender will only use strong signals if $z^0_k-z^1_k>\bi$. In this case, the attacker's utility with attack becomes $\ua=z^1_k-\beta^{\ar}_1+\bi$ if its type is $\theta^{\ar}_1$ and $\ua=z^1_k-\beta^{\ar}_2+\bi$ if its type is $\theta^{\ar}_2$, whereas without attack its utility is $z^1_k$ for both types. Thus, the attacker will choose to attack only in type $\theta^{\ar}_1$ if $\beta^{\ar}_1\leq \beta^{\dr} \leq \beta^{\ar}_2$. 
    
    Similarly, for the case of weak signals, the attacker's utility with attack becomes $\ua=z^0_k-\beta^{\ar}_1$ if its type is $\theta^{\ar}_1$ and $\ua=z^0_k-\beta^{\ar}_2$ otherwise. The attacker will then choose to attack only with type $\theta^{\ar}_1$ if $\beta^{\ar}_1\leq z^0_k-z^1_k \leq \beta^{\ar}_2$. 
    \hfill $\square$
    \end{pf}
    
     Note that in the cases specified in Proposition~\ref{sep}, the belief of the defender after it observes the attacker's actions becomes $\mu^{\ar,1}_k=1$ if the attacker attacks and $\mu^{\ar,2}_k=1$ if not.
    
    \begin{prop}\label{pool}
    In the game under Case 2b, for $n=2$, pooling equilibrium exists in the following conditions:
    \begin{itemize}
        \item Suppose $\tilde{z}_k\leq\bi$. Attacking is optimal for both types if $\beta^{\ar}_2< \bi$ and not optimal for both types if $\beta^{\ar}_1> \bi$. The optimal strategy for the defender is not to use strong signals for both types of the attacker.
        \item Suppose $\tilde{z}_k>\bi$. Attacking is optimal for both types if $\beta^{\ar}_1> \tilde{z}_k$ and not optimal for both types if $\beta^{\ar}_2< \tilde{z}_k$. The optimal strategy for the defender is to use strong signals for both types of the attacker.
    \end{itemize}
    In both cases, the defender's prior beliefs $\mu^{\dr,1}_k,$ $\mu^{\dr,2}_k$ do not change, i.e., $\mu^{\dr}_k(\theta^{\dr}_1|\ea)=\mu^{\dr,1}_k$, $\mu^{\dr}_k(\theta^{\dr}_2|\ea)=\mu^{\dr,2}_k$.
    \end{prop}
    \begin{pf}
    Again, the defender only uses strong signals if $z^0_k-z^1_k>\bi$. In this case, similar to the proof of Proposition~\ref{sep}, the attacker's utility with attack becomes $\ua=z^1_k-\beta^{\ar}_1+\bi$ if its type is $\theta^{\ar}_1$ and $\ua=z^1_k-\beta^{\ar}_2+\bi$ otherwise. Since it is known that $\beta^{\ar}_1<\beta^{\ar}_2$, given $\ua=z^1_k$ without attack, the attacker will not attack for both types if $\beta^{\ar}_1 > \bi$ and will attack if $\beta^{\ar}_1 < \bi$. 
    
    Similarly, for the case of only weak signals, since by attacking $\ua=z^0_k-\beta^{\ar}_1$ with $\theta^{\ar}_1$ type and $\ua=z^0_k-\beta^{\ar}_2$ with $\theta^{\ar}_2$ type, the attacker will not attack for both types if $\beta^{\ar}_1> z^0_k-z^1_k$ and will attack for both types if $\beta^{\ar}_2< z^0_k-z^1_k$. This completes the proof. 
    \hfill $\square$
    \end{pf}
    
    The equilibria characterized in Propositions~\ref{sep} and \ref{pool} cover all possible cases of the players' possible strategies.
    % \begin{prop}
    % For $n=2$, there is no semi-separating equilibrium.
    % \end{prop}
    % \begin{pf}
    % \textcolor{blue}{(Incomplete)}
    % \end{pf}
    
    \subsection{Effect of PBE on Consensus}
    \vspace{-0.2cm}
    We then examine how the actions of strategic players resulting from PBE affect the agents' dynamics. We note from Lemma~\ref{waste1} that the defender does not waste any of its resource by communicating weakly when there is no attack. However, as stated in the results above, there are cases where the defender will not use strong signals if the cost is too large, relative to the agent states. As a consequence, there is a case where consensus will not be achieved, especially in a signaling game setting, where the defender's knowledge is more limited.
    
    The next theorem characterizes agent consensus for the screening game setting explained in Section~\ref{41}. 

   \begin{thm}\label{inti2}
    In the game under Case 2a discussed in Section~\ref{41}, sufficient conditions to prevent consensus are $\beta^{\ar}\leq\ra$ and $\beta^{\ar}<\tilde{z}_0< \beta^{\dr}_1$.
   \end{thm}
    \begin{pf}
    From Proposition~\ref{screen}, we can see that defender does not use strong signals if $z^0_k-z^1_k<\bi$. Additionally, as long as it has enough energy, the attacker attacks if $z^0_k-z^1_k>\beta^{\ar}$. Thus, with $\ra\geq\beta^{\ar}_2$, the attacker is able to prevent consensus.   \hfill $\square$
    \end{pf}

    We further characterize agent consensus for the screening game setting explained in Section~\ref{42}.

    \begin{thm}\label{inti1}
    In the game under Case 2b, sufficient conditions for the attacker to prevent consensus are $\beta^{\ar}\leq \ra$, $\beta^{\ar}<\beta^{\dr}$, and $\tilde{z}_0\leq \beta^{\dr}$.
   \end{thm}
    \begin{pf}
    From Proposition~\ref{pool}, it is understood that the defender does not use strong signals if $z^0_k-z^1_k\leq\bi$. Additionally, the attacker attacks if $\beta^{\ar}\leq \beta^{\ar}_2<\beta^{\dr}$. Thus, given that $\ra\geq\beta^{\ar}_2\geq \beta^{\ar}$, the attacker can attack for infinite steps (including at time $k=0$ since $\ka\geq\ra$), preventing consensus.   \hfill $\square$
    \end{pf}
    
    From the conditions of Theorems~\ref{inti2}~and~\ref{inti1}, we can deduce that it is easier for the attacker to prevent consensus in signaling games since the conditions in Proposition~\ref{inti1}~are looser for the attacker. We will see through simulations if this is also the case for a more general network with $n>2$.

% %     \subsection{Comparison on Computational Complexity between BNE and PBE}
% %     Since the agents do not need to consider the attacker's action in PBE case, the agents can remove one player (at least $2^{n-1}$ possible actions) from its equilibrium computation. On the other hand, the attacker has to consider the agents' optimal response for any of its possible actions. Thus, in PBE case, the attacker needs to have much better computational resource than in BNE case, since it 
% % \vspace{-0.1cm}

\section{Numerical Examples}\label{sec5}
\vspace{-0.3cm}
% \subsection{Agents' Dynamics with BNE}
In this section, we provide numerical examples with $n>2$ to complement the results in the previous sections, especially in consensus of agents under screening and signaling games settings. Specifically, in this section we consider a path graph $1$-$2$-$3$-$4$-$5$-$6$ with $n=6$ and parameters $\ka=2$, $\ki=1.6$, $\ra=0.2$, and $\ri=0,1$. The real costs of the players are $\beta^{\ar}=0.1$ and $\beta^{\dr}=1$.

\subsection{Agents' Dynamics in Screening Games}
\begin{figure}
        \centering
        \includegraphics[trim=8 0 0 0,clip, scale=0.57]{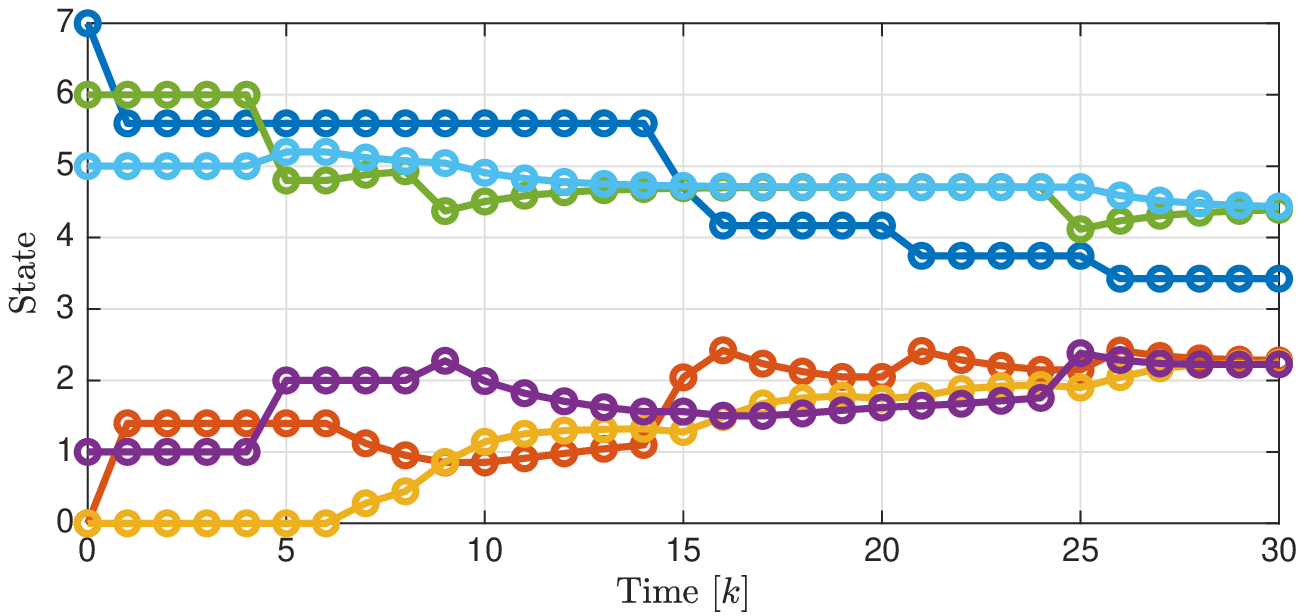}
        \vspace{-0.7cm}
        \caption{State evolution in the screening game model}
        \vspace{0.2cm}
        \label{st_scr}
        \centering
        \includegraphics[trim=8 0 0 0,clip, scale=0.57]{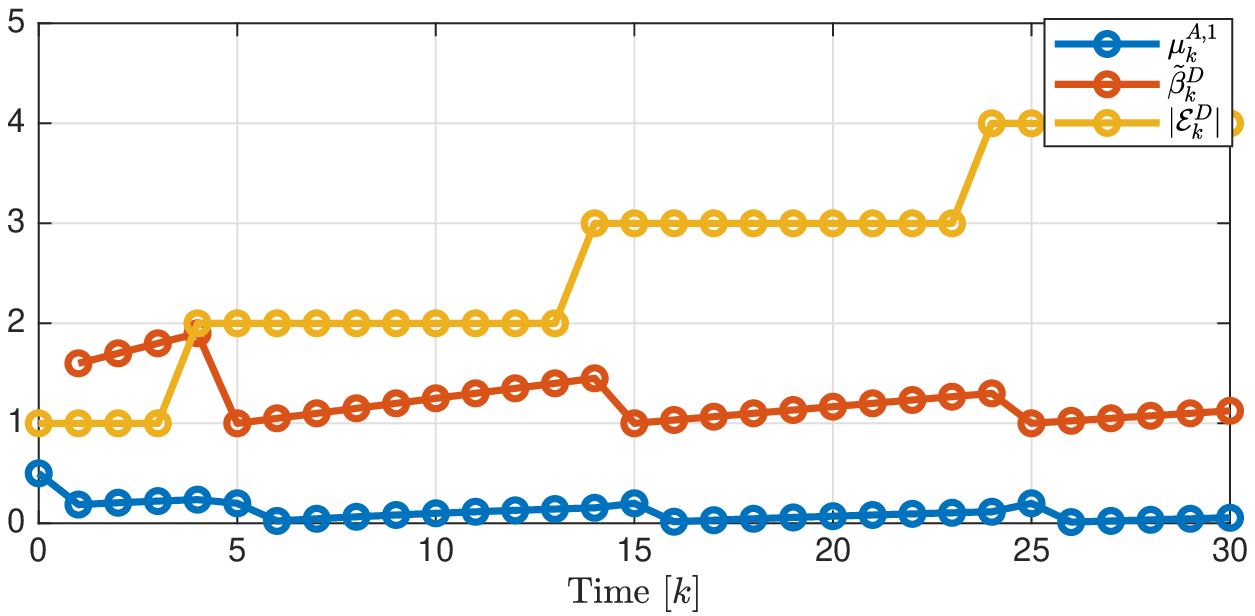}
        \vspace{-0.7cm}
        \caption{The attacker's belief $\mu^{\ar,1}_k$, the defender's estimated cost $\tilde{\beta}^{\dr}_k$, and number of edges used with strong signals $|\eo^{\dr}_k|$ in the screening games}
        \label{cost_scr}
        \vspace{0.2cm}
        \end{figure}

\vspace{-0.2cm}
First, we consider a screening game where the attacker does not know the exact value of the defender's cost $\beta^{\dr}$, with the defender's possible type $\Theta^{\dr}=\{0.5,1\}$. Fig.~\ref{st_scr} shows the evolution of the agent states $x[k]$ over time, whereas Fig.~\ref{cost_scr} illustrates the changes of the attacker's belief $\mu^{\ar,1}_k$, the defender's estimated cost $\tilde{\beta}^{\dr}_k$, and the number of edges used with strong signals $|\eo_k^{\dr}|$ over time.

From Figs.~\ref{st_scr}~and~\ref{cost_scr}, we observe that the attacker's belief of the defender's type being $\beta^{\dr}_1$ is getting closer to zero (and the estimated cost is closer to 1) every time the defender recovers (shown with increasing $|\eo^{\dr}_k|$). On the other hand, the attacker's belief $\mu^{\ar,1}_k$ slowly increases when the defender does not use strong signals. In this setting, the agent states keep getting closer despite $\rho^{\ar}>\beta^{\ar}$.

\subsection{Agents' Dynamics in Signaling Games}
\begin{figure}
        \centering
        \includegraphics[trim=8 0 0 0,clip, scale=0.57]{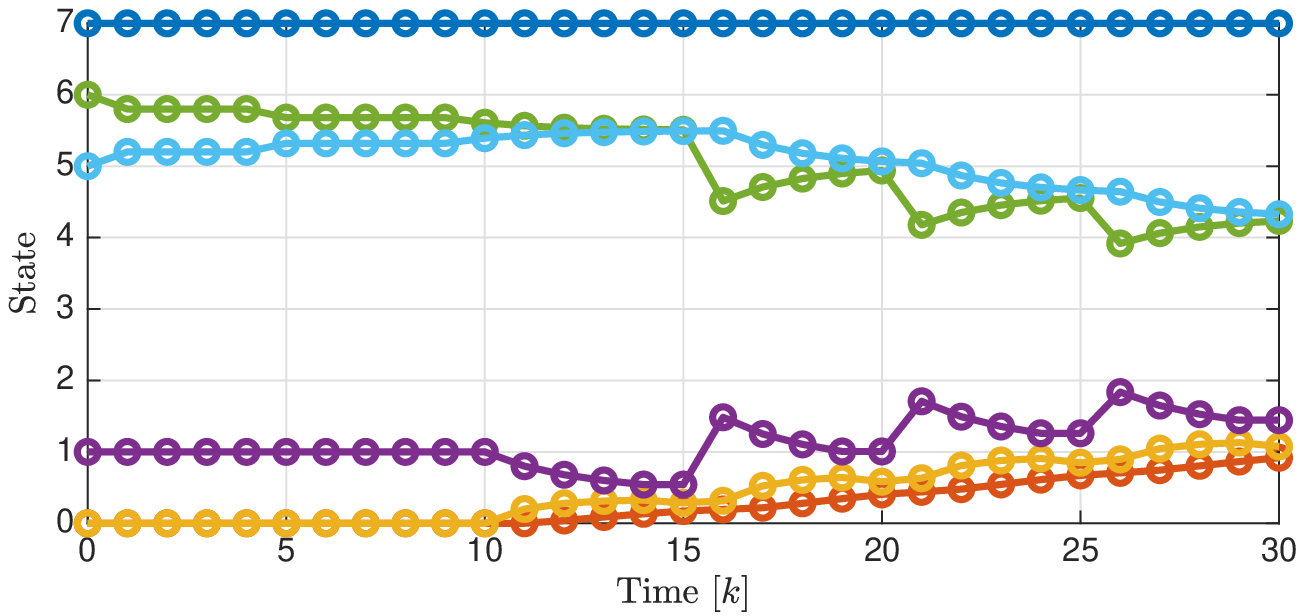}
        \vspace{-0.7cm}
        \caption{State evolution in the signaling game model}
        \vspace{0.2cm}
        \label{st_sig}
        \centering
        \includegraphics[trim=8 0 0 0,clip, scale=0.57]{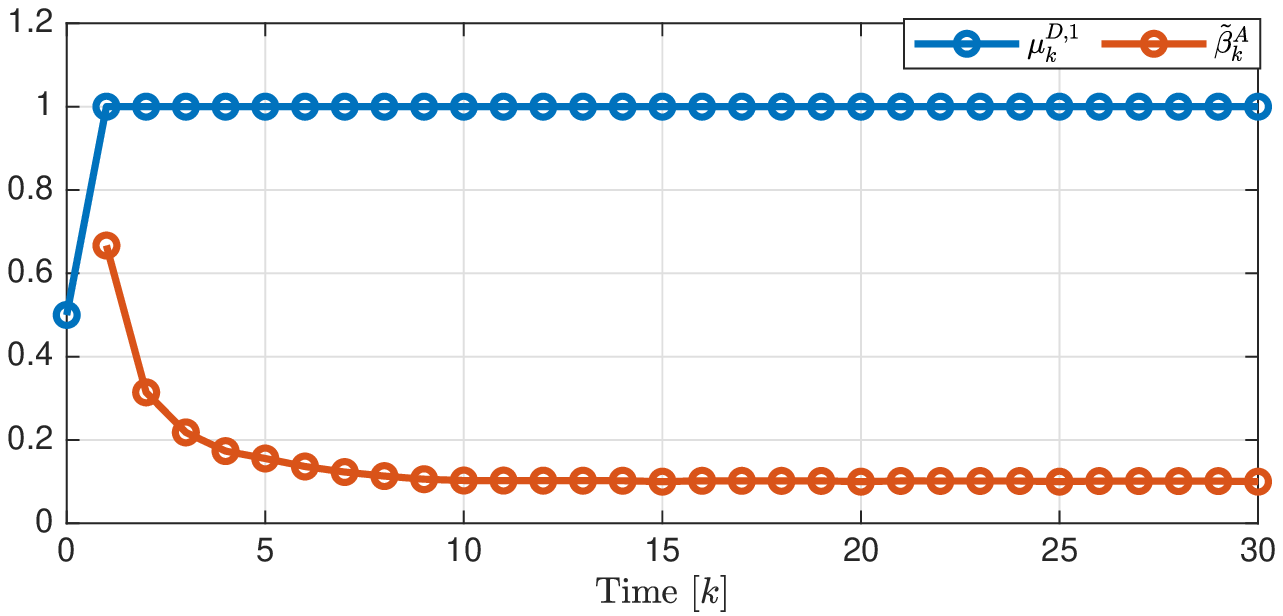}
        \vspace{-0.7cm}
        \caption{The defender's belief $\mu^{\dr,1}_k$ and the attacker's estimated cost $\tilde{\beta}^{\ar}_k$ in the signaling games}
        \vspace{0.2cm}
        \label{cost_sig}
\end{figure}

\vspace{-0.2cm}
We next consider a signaling game setting where the defender observes the attacker's action at the same time step but does not know the attacker's type as discussed in Section~\ref{thrf}, with $\Theta^{\ar}=\{0.1,1\}$. Figs.~\ref{st_sig} and~\ref{cost_sig} show the evolution of agent states over time and the changes of $\mu^{\dr,1}_k$ and $\tilde{\beta}^{\ar}_k$, respectively.

From the two figures, we notice that the agent states $x_i[k]$ are further from each other compared to the screening game case despite using the same parameters. This is inline with the statement in Section~\ref{thrf} where it is easier for the attacker to prevent consensus in the signaling game setting. The players arrive at a pooling equilibrium only for time $k=0$ and then quickly change to a separating equilibrium at $k=1$. The defender's belief $\mu^{\dr,1}_1=1$ implying that it is sure that the attacker has a low cost $\ba=0.1$ as soon as the predicted cost $\tilde{\beta}^{\ar}_k$ falls below $\beta^{\ar}_2$, as defined in (\ref{bel_d}).

\vspace{-0.3cm}

\section{Conclusion}\label{sec6}
\vspace{-0.3cm}
In this paper, we have discussed a two-player game-theoretical model of agents' communication in networks with consensus protocol under jamming attacks where players possess incomplete information of their opponents. Several game models and structures have been considered, including imperfect and perfect players' knowledge of their opponents' actions. The equilibrium of each model as well as its impact on the agents' consensus have been discussed, where the attacker is more likely to prevent consensus with complete knowledge of the opponent.

\bibliography{ifac22}

\end{document}